\documentclass[pre,twocolumn,showpacs,amsmath,amssymb]{revtex4}

\usepackage{graphicx}
\usepackage{dcolumn}
\usepackage{bm}
\usepackage{verbatim}
\usepackage{subfigure}
\usepackage{dcolumn}
\usepackage{xspace}

\newcommand{\md}{monomer-dimer\xspace}

\begin{document}

\title{Logarithmic corrections in the free energy of
  \md model on plane lattices with free boundaries}

\author{Yong Kong}
\email{matky@nus.edu.sg}
\affiliation{%
Department of Mathematics\\
National University of Singapore\\
Singapore 117543\\
}%

\date{\today}

\begin{abstract}
Using exact computations we study the 
classical hard-core \md models 
on $m \times n$ plane lattice strips with free boundaries.
For an arbitrary number $v$ of monomers (or vacancies), 
we found a 
logarithmic correction term in the finite-size correction of the
free energy.
The coefficient of the logarithmic correction term
depends on the number of monomers present ($v$) and the
parity of the width $n$ of the lattice strip:
the coefficient equals to $v$ when $n$ is odd, 
and $v/2$ when $n$ is even. The results are generalizations of the previous
results for a single monomer in an otherwise fully packed lattice of dimers.
\end{abstract}

\pacs{05.50.+q, 02.10.De, 02.70.-c, 11.25.Hf}


\maketitle

The \md problem is a classical model in statistical physics.
In the model the diatomic molecules are modeled as rigid dimers 
which occupies two adjacent sites in a regular lattice and no lattice site 
is covered by more than one dimer.  
The lattice sites that are not covered by the dimers
are regarded as occupied by monomers. 
A central problem of the model is to enumerate the 
dimer configurations on the lattice.  
A breakthrough came in 1961 when an elegant exact closed-form solution
was found for a special case 
of the model in plane lattices, namely
when the lattice is completely covered by dimers 
(the close-packed dimer problem, or dimer-covering problem) 
\cite{Kasteleyn1961}.
The method used to achieve this solution, however,
cannot be extended to tackle the more general  
\md problem where there are vacancies in the lattice,
and the problem remains unsolved notwithstanding  years of efforts.
Rigorous results exists only for 
series expansion in low dimer density 
\cite{Nagle1966},
lower bounds on free energy 
\cite{Bondy1966},
monomer-monomer correlation function of two monomers
in a lattice otherwise packed with dimers 
\cite{Fisher1963},
and locations of zeros of partition functions
\cite{Heilmann1972}.  
Some approximate methods have been proposed \cite{KenyonRS96}.
One recent advance is an analytic solution to the 
special case of the problem where there is a single 
vacancy at certain specific sites on the boundary 
of the lattice \cite{Tzeng2003}.

There is renewed interest in \md problem recently.
The interest comes from different
directions. Besides the intrinsic interest of the problem itself
and its close relation to the well-studied 
Ising model 
\cite{Kasteleyn1963},
the model also acts as the classical limit of the
recently introduced quantum dimer model 
\cite{RokhsarK88},
which has been investigated intensively as the central model
in modern theories of strongly correlated quantum matter.

The problem also attracts attention in the field of
computational complexity.  It has been shown that 
two-dimensional \md problem is 
computationally intractable and belongs to 
the \emph{``\#P-complete''} class 
\cite{Jerrum1987}.
\#P-complete class
plays the same role for counting problems 
(such as counting dimer configurations discussed here)
as the more familiar \emph{NP-complete} class for the
decision problems (such as the well-known traveling salesman problem).
The \#P-complete problems are at least as hard as the NP-complete
problems \cite{Garey1979}.
If \emph{any} problem in the \#P-complete class 
is found to be solvable, every problem in \#P class
is solvable.
Currently it is not clear whether there exists any such solution
to the \#P-complete or NP-complete class problems, and ``P verse NP'' problem
is the perhaps the major
outstanding problem in theoretical computer science.

In this Letter we report the logarithmic correction in
the free energy of the \md model with arbitrary number of monomers
on plane lattice strips of size $m \times n$, where
the width $n$ is fixed, and address the finding in the
context of universality, scaling, and exact finite-size corrections in
critical systems.
These topics 
have been studied intensively in recent
decades. 
As one of the few non-trivial exactly solved statistical lattice models,
the dimer model attracts much attention recently to test
the predictions of conformal field theory 
\cite{Blote1986}
and finite-size scaling 
\cite{Cardy1988}.
Unlike other models, the (close-packed) dimer model on plane lattice
shows certain peculiarity with respect to the predictions of the
theories, both for the central charge 
\cite{Brankov1995,Izmailian2005}
and for the logarithmic corrections \cite{Brankov1995}.
The free energy of a two-dimensional system at criticality
is assumed to have the form of 
\cite{Blote1986,Cardy1988}
\begin{equation} \label{E:theory}
 F_{m,n} = mn f_b + (m+n) f_s + C \ln{m} + D + o(1).
\end{equation}
The first two terms, $f_b$ and $f_s$,
are the ``bulk'' and ``surface'' terms, respectively. 
These two terms are not universal and depend on
the details of the model. 
The other two coefficients, $C$ and $D$,
are supposed to be universal, depending only on the shape of the system
and the boundary conditions.
The dimensionless coefficient $D$ is related to the central charge of 
the system and the boundary conditions in the transversal 
direction
\cite{Blote1986}.
The coefficient of the logarithmic term $C$ depends on the geometry
of the system. For systems with corners, as for the lattices
with free boundaries discussed here,
$
 C = \sum_i u_i, 
$ 
with some universal contribution $u_i$ from each corner
\cite{Cardy1988}. 

From the exact solution of the close-packed dimer problem,
the asymptotic expression of the free energy
$F_{m,n} = \ln a_{mn/2}$
was found to be
\cite{Ferdinand1967}
\begin{equation} \label{E:asymp}
F_{m,n}
= mn f_b + (m+n) f_s + D_n[(m+1)/(n+1)] + o(n^{-2+\delta})
\end{equation}
where $\delta > 0$ and $D_n$ depends on the geometry of the lattice and the
parity of $n$. Two issues arise from this asymptotic expression.
The first issue is related to the central charge, because
the expansion depends on the parity of $n$:
\[
 \frac{F_{m,n}}{mn} = \frac{G}{\pi} 
     + [\frac{G}{\pi} - \frac{1}{2} \ln(1+\sqrt{2})] \frac{1}{n} 
     + \frac{\pi}{24} \frac{1}{n^2} + \cdots
\]
when $n$ is even, and
\[
 \frac{F_{m,n}}{mn} = \frac{G}{\pi} 
     + [\frac{G}{\pi} - \frac{1}{2} \ln(1+\sqrt{2})] \frac{1}{n} 
     - \frac{\pi}{12} \frac{1}{n^2} + \cdots
\]
when $n$ is odd. Here $G$ is the Catalan's constant.
At the first look, the coefficients of $n^{-2}$, which are
related to the central charge of the conformal field theory,
are different for even and odd $n$, leading to
$c=1$ for the even $n$ and $c=-2$ for odd $n$.
Recently this issue has been discussed using logarithmic 
conformal field theory \cite{Izmailian2005,Itzykson1986}. 
It is pointed out that the coefficients of $n^{-2}$ are actually
related to the \emph{effective} central charge
$c_{\text{eff}} = c - 24 h_{\min}$, where $h_{\min}$
is the conformal weight of the ground state 
and is a boundary dependent quantity. 
By using a bijection of the dimer coverings
with a spanning tree and Abelian sandpile model,
it is found that changing the parity of $n$ 
results in the change of the boundary conditions of the 
mapped spanning trees, leading to different $c_{\text{eff}}$.
The central charge itself, however, remains unchanged $c = -2$
\cite{Izmailian2005}.

The second issue,
which was pointed out previously \cite{Brankov1995},
is the absence of corner contribution term
predicted by the theory (Eq. \ref{E:theory})
in the expansion of Eq. (\ref{E:asymp}) of the close-packed dimer
model.
This issue, however, changes its nature when we
look at the general \md model.
Recently we found that when there is one monomer in the
lattices, there is indeed a logarithmic term in the finite-size
correction of the free energy \cite{Kong2006}.
In the following we generalize the results to plane lattices with 
arbitrary number of monomers, and found that the coefficients
of the logarithmic correction term depend not only on the
number of monomers, but also on the parity of the width
of the lattice strip $n$.

The configurational grand 
canonical partition function of the system is
\[
Z_{m,n}(x) = a_N x^N + a_{N-1} x^{N-1} + \cdots + a_0
\]
where $a_k$ is the number of ways to arrange $k$ dimers on the
$m \times n$ plane lattice with free boundaries. 
We are interested in the coefficients
of fixed number $v$ of monomers, that is, $a_{(mn-v)/2}$,
when the width of the lattice strip $n$ is fixed and
the length of the strip $m$ changes.
When the width of the lattice strip $n$ is even, 
$v$ can only take even numbers;
when $n$ is odd, $v$ can take either even or odd values
based on the parity of $m$.

The full partition function have been calculated exactly
for $n$ from $1$ to $16$, using the method discussed 
in Refs. \cite{Kong1999, Kong2006}.
The coefficients $a_{(mn-v)/2}$ of the partition functions 
are extracted to fit the following function:
\begin{equation} \label{E:fit}
 \frac{\ln a_{(mn-v)/2}}{m n} = c_0 + \frac{c_1}{m} + \frac{c_2}{m^2} + 
 \frac{c_3}{m^3} + \frac{c_4}{m^4} + \frac{b  v  l}{n} \frac{\ln(m+1)}{m}.
\end{equation}
where $b=1$ when $n$ is odd, and $b=1/2$ when $n$ is even.
The reason to choose $\ln(m+1)$ instead of $\ln(m)$ is discussed
in Ref. \cite{Kong2006}, and the choice will not affect the results
discussed below.

\begin{figure}
  \centering
  \includegraphics[angle=270,width=\columnwidth]{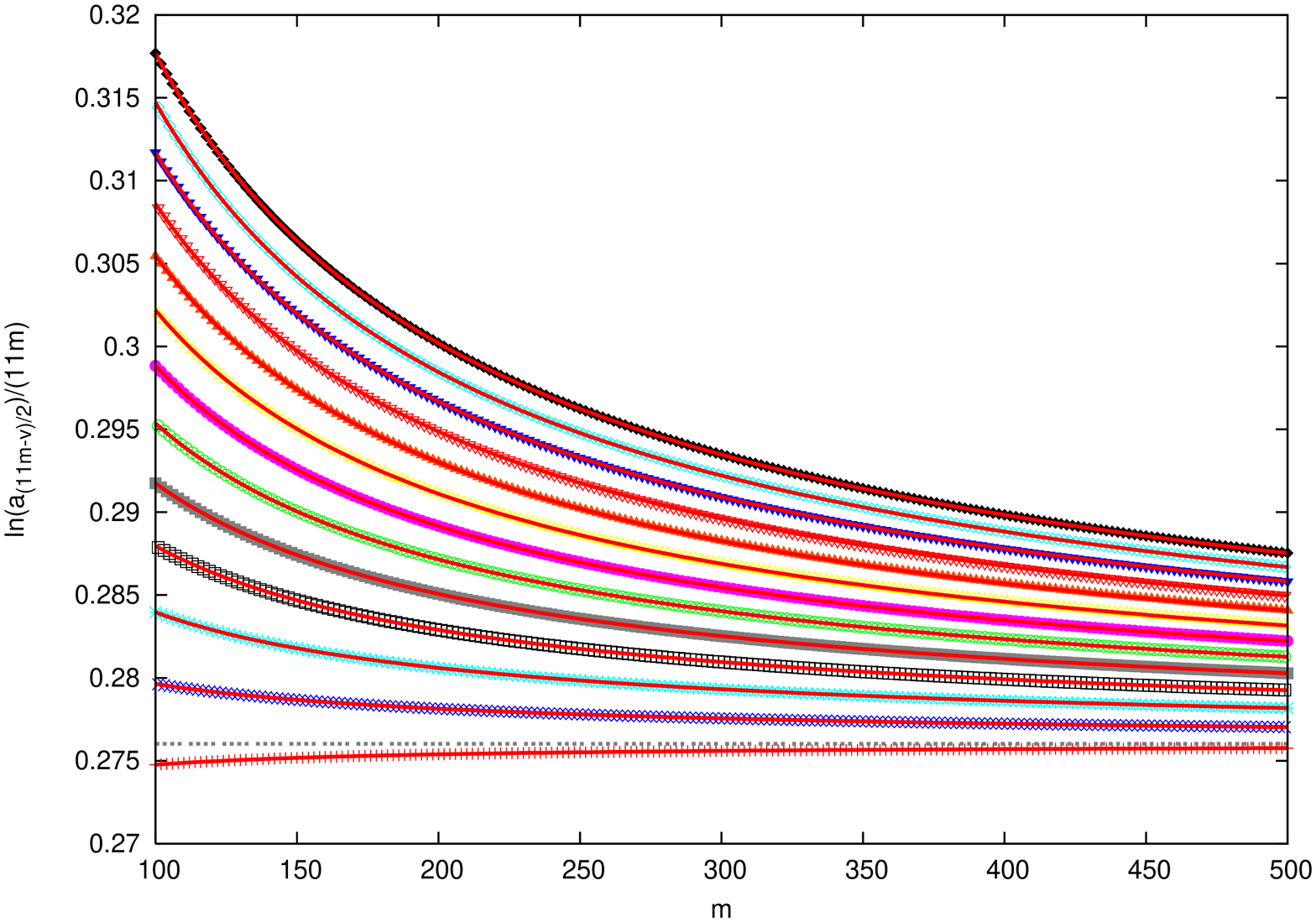}
  \caption{(Color online)
    The original data of $(mn)^{-1}\ln(a_{(mn-v)/2})$ 
    and the fitted curves for $n=11$ and $v=0, 1, \dots, 12$.
    The curve for $v=0$ is on the bottom and that
    for $v=12$ is on the top.
    The dashed horizontal line is $c_0^e(11) = 0.276016$
    from exact expression Eq. \ref{E:a_0_e}.
    \label{F:fit-11}}
\end{figure}

\begin{figure}
  \centering
  \includegraphics[angle=270,width=\columnwidth]{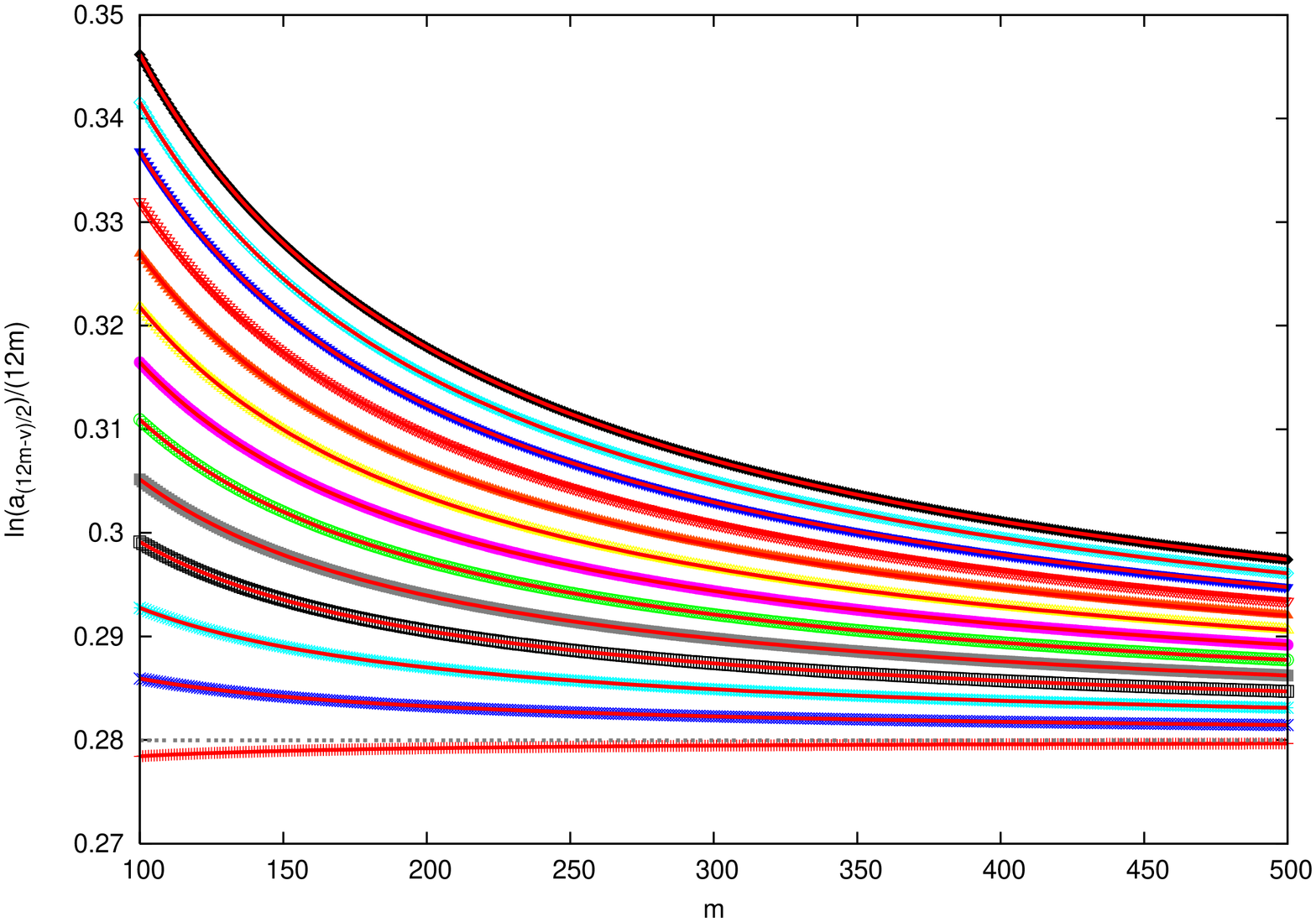}
  \caption{(Color online)
    The original data of $(mn)^{-1}\ln(a_{(mn-v)/2})$ 
    and the fitted curves for $n=12$ and $v=0, 2, \dots, 24$.
    The curve for $v=0$ is on the bottom and that
    for $v=24$ is on the top.
    The dashed horizontal line is $c_0^e(12) = 0.279976$
    from exact expression Eq. \ref{E:a_0_e}.
    \label{F:fit-12}}
\end{figure}

For the close-packed dimer problem (where $v=0$ and $mn$ is even), 
the exact solution gives $c_0$ in Eq. (\ref{E:fit}) as \cite{Kasteleyn1961}
\begin{equation} \label{E:a_0_e}
  c_0^e(n) = \frac{1}{n} \ln \left[ \prod_{l=1}^{\frac{n}{2}}
  \left( \cos \frac{l \pi}{n+1}
  + \left(1 + \cos^2 \frac{l \pi}{n+1} \right)^{\frac{1}{2}} 
  \right) \right].
\end{equation}
For a given $n$, this exact expression of $c_0$ 
actually holds for \emph{any} 
finite number $v$ of monomers. This was confirmed by fitting the data
with $c_0$ as a free parameters. In the following we use
$c_0^e(n)$ given in Eq. (\ref{E:a_0_e}) for $c_0$.
We fit the data extracted from the full partition functions 
to Eq. (\ref{E:fit}) for $v=0, \dots, 12$ for odd $n$,
and for $v=0, 2, 4, \dots, 24$ for even $n$.
The  fitting results for $l$ are shown in Tables \ref{T:fit_odd} and 
\ref{T:fit_even}, respectively \footnote{
Exact result exists for $n=1$: 
$a_{(m-v)/2} = \binom{(m+v)/2}{v}
= (2^v v!)^{-1} (m+v)(m+v-2) \cdots (m-v+2)$, which gives $l=1$ exactly.
The fitting results as listed in Table \ref{T:fit_odd} are used
as a check for the fitting procedure.
}.
As we did previously \cite{Kong2006},
in these fittings, only $m \ge m_0 = 100$ are used.
The curves of fitting for two values of $n$, $n=11$ and $n=12$,
are shown in Figure~\ref{F:fit-11} and Figure~\ref{F:fit-12}, respectively.
The fitting results for all the parameters when $n=11$ and $n=12$
are shown in Tables \ref{T:fit_11} and 
\ref{T:fit_12}, respectively.
These results lead clearly to the conclusion that
for a fixed number of monomers $v$, there is a logarithmic correction
term in the free energy $(mn)^{-1}\ln a_{(mn-v)/2}$, 
and the coefficient of this term not only
depends on $v$, but also on the parity of $n$, the width of the
lattice strip: when $n$ is odd, the coefficient equals to $v$; when
$n$ is even, the coefficient is $v/2$.
The lack of the logarithmic correction term found in the close-packed
dimer problem ($v=0$) \cite{Kasteleyn1961,Brankov1995,Ferdinand1967} 
and the logarithmic correction term 
found in the odd-by-odd lattice with one single vacancy ($v=1$) \cite{Kong2006}
are just special cases of this general result.

\begingroup
\squeezetable
\begin{table*}
\caption{Coefficient of the logarithmic term $l$ from 
fitting $ (mn)^{-1} \ln a_{(mn-v)/2}$ to Eq. \ref{E:fit} for \emph{odd} 
values of $n$, with $b=1$.
Only data with $m \ge m_0 = 100$ are used in the fitting. 
For $n=1$, $m \le 2000$; for $n=5, \dots 13$, $m \le 500$; for $n=15$,
$m \le 200$. 
Numbers in square brackets denote powers of 10.
\label{T:fit_odd}}
\begin{ruledtabular}
\begin{tabular}{ccccccccc}
      & \multicolumn{8}{c}{n} \\\cline{2-9}
 $v$  
  & \mbox{$1$} 
  & \mbox{$3$} 
  & \mbox{$5$} 
  & \mbox{$7$} 
  & \mbox{$9$} 
  & \mbox{$11$} 
  & \mbox{$13$} 
  & \mbox{$15$} 
\\\hline
0 & 
  0             &
  -1.42355[-12] &
  6.9891[-10]   &
  7.10235[-10]  &
  1.03087[-9]  & 
  1.56222[-9]  &
  3.76785[-9]  &
  3.42299[-9]\\
1 & 
   1 &
   1 &
   1 &
   1 &
   1 &
   1 &
   1 &
   1 \\
2 &
   1        &
   1        &
   1        &
   0.999999 &
   0.999998 &
   0.999997 &
   0.999993 &
   0.999968 \\
3 & 
   1        &
   0.999998 &
   0.999991 &
   0.999975 &
   0.999945 &
   0.999897 &
   0.999781 &
   0.999088 \\
4 &
   1        &
   0.999988 &
   0.999945 &
   0.999851 &
   0.99969  &
   0.999449 &
   0.998907 &
   0.996177 \\
5 & 
   1        &
   0.999962 &
   0.999835 &
   0.999576 &
   0.999163 &
   0.998585 &
   0.99736  &
   0.991937 \\
6 &
   1        &
   0.999912 &
   0.999632 &
   0.999108 &
   0.998322 &
   0.997282 &
   0.995194 &
   0.986896 \\
7 & 
   1        &
   0.999836 &
   0.999351 &
   0.998494 &
   0.997273 &
   0.995726 &
   0.99273  &
   0.981618 \\
8 & 
   1        &
   0.999721 &
   0.998951 &
   0.997671 &
   0.995927 &
   0.993799 &
   0.98982  &
   0.976008 \\
9 & 
   1        &
   0.999581 &
   0.998493 &
   0.996767 &
   0.9945   &
   0.991812 &
   0.986905 &
   0.970599 \\
10 & 
   1        &
   0.99939  &
   0.997902 &
   0.995649 &
   0.99279  &
   0.989492 &
   0.983621 &
   0.965005 \\
11 & 
   1        &
   0.999181 &
   0.997289 &
   0.994527 &
   0.991117 &
   0.987267 &
   0.980527 &
   0.959804 \\
12 &
   0.999999 &
   0.998909 &
   0.996529 &
   0.99318  &
   0.989159 &
   0.984713 &
   0.977081 &
   0.954453 \\ 
\end{tabular}
\end{ruledtabular}
\end{table*}
\endgroup

\begingroup
\squeezetable
\begin{table*}
\caption{Coefficient of the logarithmic term $l$ from
fitting $(mn)^{-1} \ln a_{(mn-v)/2}$ to Eq. \ref{E:fit} for \emph{even}
 values of $n$, with $b=1/2$.
Only data with $m \ge m_0 = 100$ are used in the fitting.
For $n=2$, $m \le 2000$; for $n=4$, $m \le 1000$,
for $n=6, \dots 14$, $m \le 500$; for $n=16$,
$m \le 130$.
Numbers in square brackets denote powers of 10.
\label{T:fit_even}}
\begin{ruledtabular}
\begin{tabular}{ccccccccc}
      & \multicolumn{8}{c}{n} \\\cline{2-9}
 $v$
 & \mbox{$2$} 
 & \mbox{$4$} 
 & \mbox{$6$} 
 & \mbox{$8$} 
 & \mbox{$10$} 
 & \mbox{$12$} 
 & \mbox{$14$} 
 & \mbox{$16$} 
\\\hline
0 &
  1.21643[-11] &
  1.3329[-10]  &
  5.89405[-10] &
  9.28482[-10] &
  9.0419[-10]  &
  1.13781[-9] &
  8.97751[-8] &
  9.45733[-5] \\
2 &
  1 &
  1 & 
  1 & 
  0.999999 &
  0.999999 &
  1        & 
  1.00017  &
  1.00218  \\
4 &
  1       &
  1       &
  1.00001 &
  1.00002 &
  1.00006 &
  1.00012 &
  1.00075 &
  1.00026 \\
6 &
  1       &
  1       &
  1.00002 &
  1.00003 &
  1.00004 &
  1       &
  0.998332 &
  0.990178 \\
8 &
  1        &
  1        &
  0.999994 &
  0.999931 &
  0.999739 &
  0.999318 &
  0.993311 &
  0.98074  \\
10 &
  1        &
  0.999997 &
  0.999907 &
  0.999639 &
  0.999043 &
  0.998001 &
  0.987848 &
  0.977344 \\
12 &
  1        &
  0.999982 &
  0.999718 &
  0.999103 &
  0.997943 &
  0.996201 &
  0.983945 &
  0.981147 \\
14 &
  1        &
  0.999951 &
  0.9994   &
  0.998314 &
  0.996526 &
  0.994189 &
  0.982769 &
  0.991167 \\
16 &
  1        &
  0.999897 &
  0.998938 &
  0.997305 &
  0.994942 &
  0.99226  &
  0.984712 &
  1.00579 \\
18 &
  0.999997 &
  0.999816 &
  0.998332 & 
  0.996139 &
  0.993357 &
  0.990678 &
  0.989654 &
  1.02348 \\
20 &
  0.999992 &
  0.999702 &
  0.997596 &
  0.994896 &
  0.991933 &
  0.989648 &
  0.997209 &
  1.04299 \\
22 &
  0.999983 &
  0.999551 &
  0.996756 &
  0.993664 &
  0.990808 &
  0.989309 &
  1.00689  &
  1.06337 \\
24 &
  0.99997  &
  0.999361 &
  0.995843 &
  0.992527 &
  0.990091 &
  0.989743 &
  1.01822  &
  1.08398  
\end{tabular}
\end{ruledtabular}
\end{table*}
\endgroup

\begingroup
\squeezetable
\begin{table}
\caption{Fitting $(mn)^{-1} \ln a_{(mn-v)/2}$ to Eq. \ref{E:fit} 
 for $n=11$ with $b=1$ and 
 $c_0$ fixed as $c_0^e(11) = 0.276016$ given by Eq. \ref{E:a_0_e}.
Only data with $m \ge m_0 = 100$ are used in the fitting.
Numbers in square brackets denote powers of 10.
\label{T:fit_11}}
\begin{ruledtabular}
\begin{tabular}{ccccccc}
 $v$  
  & \mbox{$c_1$} & \mbox{$c_2$} & \mbox{$c_3$} 
  & \mbox{$c_4$} & \mbox{$l$}\\\hline
0 & 
    -0.125966   & 8.75546[-8] & -8.24421[-6] & 0.000319302 &
    1.56222[-9] \\
1 & 
    -0.0561077  & -0.140372   & 0.0319911    & -0.0345653  &
    1 \\
2 & 
    -0.0492583  & 0.474695    & -4.7249      & 23.4862     &
    0.999997 \\
3 & 
    -0.0790665  & 1.82165     & -36.8528     & 449.079     &
    0.999897 \\
4 & 
    -0.133937   & 3.79131     & -103.787     & 1715.28     &
    0.999449 \\
5 & 
    -0.207035   & 6.24906     & -203.292     & 3861.49     &
    0.998585 \\
6 & 
    -0.293685   & 9.05639     & -328.77      & 6728.3      &
    0.997282 \\
7 & 
    -0.391277   & 12.162      & -480.3       & 10421.5     &
    0.995726 \\
8 & 
    -0.496769   & 15.4114     & -645.37      & 14479.9     &
    0.993799 \\
9 & 
    -0.609818   & 18.8669     & -833.317     & 19374.1     &
    0.991812 \\
10 & 
    -0.727328   & 22.305      & -1022.43     & 24212.7     &
    0.989492 \\
11 & 
    -0.850844   & 25.9246     & -1235.21     & 30014.7     &
    0.987267 \\
12 & 
    -0.976389   & 29.3893     & -1437.02     & 35307.6     &
    0.984713
\end{tabular}
\end{ruledtabular}
\end{table}
\endgroup

\begingroup
\squeezetable
\begin{table}
\caption{Fitting $(mn)^{-1} \ln a_{(mn-v)/2}$
  to Eq. \ref{E:fit} for $n=12$ with $b=1/2$ and 
 $c_0$ fixed as $c_0^e(12) = 0.279976$ given by Eq. \ref{E:a_0_e}.
Only data with $m \ge m_0 = 100$ are used in the fitting.
Numbers in square brackets denote powers of 10.
\label{T:fit_12}}
\begin{ruledtabular}
\begin{tabular}{ccccccc}
 $v$  
  & \mbox{$c_1$} & \mbox{$c_2$} & \mbox{$c_3$} 
  & \mbox{$c_4$} & \mbox{$l$}\\\hline
0 & 
    -0.154113   &  5.91813[-8]   &   -5.64565[-6]   & 0.000221538 &
    1.13781[-9] \\
2 & 
    0.219858    &  -0.748376     &   -2.59592       & -16.0843    &
    1 \\
4 & 
    0.535917    &  -2.62009      &   -9.74674       & 325.144     &
    1.00012 \\
6 & 
    0.818481    &  -5.66056      &   -12.3135       & 1291.39     &
    1 \\
8 & 
    1.07862     &  -10.0074      &   12.3026        & 2471.16     &
    0.999318 \\
10 & 
    1.32299     &  -15.8166      &   92.4989        & 2982.18     &
    0.998001 \\
12 & 
    1.55565     &  -23.1898      &   253.569        & 1852.54     &
    0.996201 \\
14 & 
    1.77871     &  -32.1427      &   512.746        & -1715.1     &
    0.994189 \\
16 & 
    1.99286     &  -42.6063      &   878.094        & -8235.31    &
    0.99226  \\
18 & 
    2.19782     &  -54.4469      &   1349.67        & -17937.8    &
    0.990678 \\
20 & 
    2.39273     &  -67.4891      &   1921.53        & -30816.5    &
    0.989648 \\
22 & 
    2.57645     &  -81.5376      &   2583.82        & -46694.5    &
    0.989309 \\
24 & 
    2.74778     &  -96.3937      &   3324.52        & -65284.5    &
    0.989743
\end{tabular}
\end{ruledtabular}
\end{table}
\endgroup

The close-packed dimer model, which is a special case of the general
\md model discussed here, already demonstrates intriguing connections
with the theory of university and finite scaling 
\cite{Brankov1995,Izmailian2005}.
It would be interesting to see how the results reported
here incorporate into the bigger picture.


\end{document}